\documentclass[fleqn,12pt,twoside]{article}
\usepackage[headings]{espcrc1}
\usepackage[english]{babel}

\usepackage{amsfonts}
\usepackage{amssymb}
\usepackage{amsmath}

\usepackage[dvips]{graphicx}

\usepackage{latexsym,epsf,axodraw,pstricks}
\usepackage{boxedminipage,floatflt,fancybox}

\usepackage{amscd}
\usepackage{psfrag}

\newcommand{\be}{\begin{equation}}
\newcommand{\ee}{\end{equation}}
\newcommand{\bea}{\begin{eqnarray}}
\newcommand{\eea}{\end{eqnarray}}
\newcommand{\bse}{\begin{subequations}}
\newcommand{\ese}{\end{subequations}}
\newcommand{\ba}{\begin{align}}
\newcommand{\ea}{\end{align}}

\newcommand{\bc}{\begin{center}}
\newcommand{\ec}{\end{center}}
\newcommand{\bi}{\begin{itemize}}
\newcommand{\ei}{\end{itemize}}

\newcommand{\bfig}{\begin{figure}[h]}
\newcommand{\efig}{\end{figure}}
\newcommand{\incfig}{\includegraphics}

\newcommand{\Gs}{{\bar{G}}}
\newcommand{\phis}{{\bar{\phi}}}

\newcommand{\half}{\frac{1}{2}}

\newcommand{\ts}[1]{{\mbox{\scriptsize #1}}}

\newcommand{\tpic}[1]{\;\parbox[c]{20pt}{\begin{picture}(20,30)(0,0)
\SetWidth{1.0}\SetScale{1.0} #1 \end{picture}}\;}
\newcommand{\pic}[1]{\;\parbox[c]{30pt}{\begin{picture}(30,30)(0,0)
\SetWidth{1.0}\SetScale{1.0} #1 \end{picture}}\;}

\newcommand{\picb}[1]{\;\parbox[c]{45pt}{\begin{picture}(45,30)(0,0)
\SetWidth{1.0}\SetScale{1.0} #1 \end{picture}}\;}

\title{Renormalized Finite Temperature $\mathbf{\phi^4}$ theory
from the 2PI Effective Action}

\author{A. Arrizabalaga\address{NIKHEF \\ Kruislaan 409, 1098 SJ, Amsterdam, The Netherlands}
		and
		U. Reinosa\address{Institut f\"ur Theoretische Physik, Universit\"at Heidelberg,\\
							Philosophenweg 16, 69120 Heidelberg, Germany}.}

\begin{document}

\maketitle

\begin{abstract}
We present a numerical and analytical study of scalar $\phi^4$ theory at finite temperature with a renormalized 2-loop truncation of
the 2PI effective action.
\end{abstract}

\section{Introduction}
The thermodynamical properties of a physical system can be extracted from the effective potential, which is given
as a function $\gamma[\phi]$ of a condensate field $\phi$. In the thermodynamic limit, its stationary
point defines the free energy $\mathcal{F}$ as
\begin{equation}
\mathcal{F}=\lim_{V\rightarrow \infty}\gamma[\bar{\phi}],\quad\mbox{with}\quad \frac{\delta \gamma[\phi]}{\delta
\phi}\Big|_{\phi=\bar\phi}=0.
\end{equation}
Once the free energy is known, all other thermodynamic quantities, such as the entropy or energy density, can be
easily derived. At finite temperature collective phenomena are known to modify substantially the properties of the
elementary excitations, so the use of a perturbative expansion around the free theory to calculate the effective
potential becomes questionable and often leads to inconsistencies. One needs then to consider nonperturbative schemes.
For situations where the effects of the collective phenomena can be conveniently captured by a suitable modification of
the 2-point functions, resummation schemes based on the 2PI
effective action can be very useful \cite{2PIequilibrium}.  The 2PI effective action provides a reorganization of the
perturbative expansion around dressed 2-point functions, which are determined self-consistently for a given
approximation/truncation.
In this work we study the thermodynamics of scalar $\phi^4$ theory using a 2-loop
truncation of the 2PI effective action.

\section{2-loop truncation of the 2PI effective action}
For scalar $\phi^4$ theory the 2PI effective action can be parametrized as \cite{CJT}
\begin{equation}
\Gamma_\ts{2PI}[\phi,G]=\half \phi\cdot G_0^{-1}\cdot \phi+\half \mbox{Tr}\big[\mbox{ln} G^{-1}+\left(G_0^{-1}-G^{-1}\right)\cdot
G\big]+\Gamma_{\ts{int}}[\phi,G],
\end{equation}
with $\phi$ and $G$ generic 1- and 2-point functions and $A\cdot B$ a shorthand notation for the convolution of $A$ and $B$. The term
$\Gamma_{\ts{int}}$ contains the interactions and can be written as an expansion in terms of two-particle-irreducible
(2PI) diagrams. Up to 2-loops it is given by
\footnote{The Feynman rules are given by: $\tpic{
\Line(5,5)(25,25)
\Line(5,25)(25,5)
\Vertex(15,15){2}
}=-\lambda\,, \qquad 
\pic{
\Line(0,15)(30,15)
}=G
\,, \qquad 
\pic{
\Line(5,15)(23,15)
\GCirc(23,15){2.5}{1}
\Line(25,17)(21,13)
\Line(21,17)(25,13)
}=\phi
\,.
$}
\be
-\Gamma_\ts{int}[\phi,G]=
\frac{1}{4!}
\tpic{
\Vertex(10,15){2}
\GCirc(2,7){2.5}{1}
\GCirc(18,7){2.5}{1}
\Line(10,15)(0,5)
\Line(10,15)(20,5)
\Line(0,9)(4,5)
\Line(20,9)(16,5)
\GCirc(2,23){2.5}{1}
\GCirc(18,23){2.5}{1}
\Line(10,15)(0,25)
\Line(10,15)(20,25)
\Line(0,21)(4,25)
\Line(20,21)(16,25)
}
+	
\frac{1}{4}
\tpic{
\Oval(10,22.5)(7.5,7.5)(0)
\Vertex(10,15){2}
\GCirc(2,7){2.5}{1}
\GCirc(18,7){2.5}{1}
\Line(10,15)(0,5)
\Line(10,15)(20,5)
\Line(0,9)(4,5)
\Line(20,9)(16,5)
}
+
\frac{1}{8}
\tpic{
\Oval(15,7.5)(7.5,7.5)(0)
\Oval(15,22.5)(7.5,7.5)(0)
\Vertex(15,15){2}
}
+\frac{1}{12}
\picb{
\GCirc(22.5,15){11.5}{1}
\Line(2,15)(43,15)
\Vertex(11,15){2}
\Vertex(34,15){2}
\GCirc(2,15){2.5}{1}
\GCirc(43,15){2.5}{1}
\Line(0,17)(4,13)
\Line(0,13)(4,17)
\Line(45,17)(41,13)
\Line(45,13)(41,17)
}
\ee
Within the approximation, ``physical'' 1- and 2-point functions $\bar\phi$ and $\bar{G}$ are determined self-consistenly
as the stationary points of the
2PI effective action, i.e.
\be
\frac{\delta \Gamma_\ts{2PI}}{\delta G}\Big|_{\bar{G}[\phi],\phi}=0\quad\mbox{and}\quad \frac{\delta \Gamma_\ts{2PI}}{\delta
\phi}\Big|_{\bar{G}[\phi],\bar\phi}=0.
\label{eq:stationarity}
\ee
The stationary conditions (\ref{eq:stationarity}) turn into a set of coupled implicit equations for $\Gs$ and
$\bar\phi$ which have to be solved in order to calculate any physical quantity. In particular, the knowledge of the dressed 2-point function $\bar{G}$ allows one to calculate the effective potential
as $\gamma[\phi]=TV^{-1}\Gamma_{\ts{2PI}}[\phi,\bar{G}[\phi]]$. 
\par
One of the main complications that arises when dealing with truncations of the 2PI effective action is the fact that 2-
and higher $n$-point functions are \emph{not uniquely defined} \cite{ourpaper}. In particular, a given truncation
defines two possible 2-point functions
and three 4-point functions. One of the 2-point functions is given by the stationary value $\bar{G}$, while the other is
related to the inverse curvature of the effective potential as $\hat{G}^{-1}=TV^{-1}\left[\delta^2 \gamma[\phi]/\delta
\phi^2\right]$. The three 4-point functions and their
corresponding vertices are given in terms of coupled Bethe-Salpeter-like equations \cite{ourpaper}.
Concerning \emph{renormalization}, the ambiguity in the definition of the vertex functions implies that there are more than
one counterterm of a given type \cite{renormalization}. A given counterterm is determined by applying a renormalization condition to the
corresponding vertex. For consistency, identical renormalization conditions are applied to all counterterms of the same
type.  It is simpler to consider renormalization conditions applied at a reference temperature $T_\star$ for which the
physical field configuration is $\bar\phi_\star=0$. This requires that $T_\star>T_c$ if a critical temperature $T_c$
exists. 
Defining the self-energy from Dyson's equation $\bar\Sigma=\Gs^{-1}-G_0^{-1}$, the renormalized equations for $\bar{\Sigma}$ and
$\bar\phi$ are
\begin{align}
\bar\Sigma(P)&=\delta m_0^2+\frac{\lambda+\delta
\lambda_2}{2}\phi^2+\frac{\lambda+\delta\lambda_0}{2}\int_K^T \bar{G}(K)-\frac{\lambda^2}{2}\phi^2\Theta(P)
\label{eq:gap}\\
\left(m^2+\delta m_2^2\right)\phis&=\frac{\lambda+\delta \lambda_4}{6}\phis^3+\frac{\lambda+\delta
\lambda_2}{2}\phis\int_K^{T}\Gs(K)-\frac{\lambda^2}{6}\phis\int_K^T \Gs(K)\Theta(K), \label{eq:field}
\end{align}
with $\smash{\int_K^T}$ a short-hand notation for the standard sum-integral over momentum $K$ at temperature $T$, and
$\smash{\Theta(P)=\int_K^T \Gs(K)\Gs(K+P)}$. The bar in $\phi$ has been omitted in the first equation to stress the fact that it
can be solved for any value. This is needed, for instance, to calculate the effective potential $\gamma[\phi]$.
The explicit expressions for the counterterms $\delta m_0^2$, $\delta m_2^2$,
$\delta \lambda_0$, $\delta \lambda_2$ and $\delta \lambda_4$ can be found in Ref.~\cite{ourpaper}. With those
counterterms it can be shown
that, for any value of $T$ and $\phi$, the results for $\Gs$, $\phis$ and $\gamma[\phi]$ are UV-finite.

\section{Numerical Analysis}

\begin{figure}
\vspace{-.25cm}
\centerline{\incfig[width=.45\textwidth]{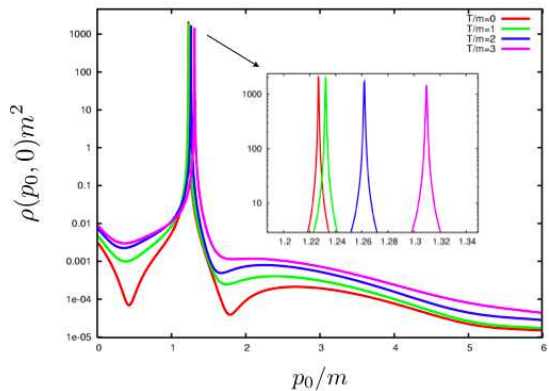}}
\vspace{-1cm}
\label{fig:spectral}
\caption{Spectral function $\rho(p_0,0)$ for $\lambda=1$, $\phi/m=1$ and $T_\star/m=1$ and several
temperatures.}
\vspace{-.75cm}
\end{figure}

The gap and field equations (\ref{eq:gap}) and (\ref{eq:field}) are solved numerically in Minkowski space. We split the self-energy into a local and a non-local part as
$\bar\Sigma(P)=\bar\Sigma^l+\bar\Sigma^{nl}(P)$. The momentum dependent part is expanded in both $p_0$ and $|\mathbf{p}|$ using $N$ Chebyshev polynomials as
$\smash{\bar\Sigma^{nl}(p_0,|\mathbf{p}|)=\sum_i^N \sum_j^N c_{ij}T_i(p_0)T_j(|\mathbf{p}|)}$.
The solution to the gap and field equations is obtained by solving the matrix equation for the $N^2$ coefficients
$c_{ij}$ plus the local terms $\Sigma^l$ and/or $\phis^2$ by means of a multi-dimensional Newton-Raphson method.
This numerical algorithm is fairly stable and converges after few iterations. For a given $N$, the numerical
solution oscillates around the ``full'' solution, which is convenient for the calculation of integrated quantities, such as
the effective potential. Unlike lattice-based methods, this technique also allows the use of large (small) UV (IR) cut-offs,
which is useful for the study of \emph{renormalization} and/or \emph{critical phenomena}.
Although numerically more expensive, the advantage of solving the equations in Minkowski space is that one can obtain directly spectral properties of the
system without the need of analytic continuation. In particular, the spectral function $\rho(p_0,\mathbf{|p|})$ can be constructed from the knowledge of the real and imaginary parts of the retarded
self-energy, which come directly from solving equation (\ref{eq:gap}) (see Fig.~\ref{fig:spectral} for an example result
at several
temperatures). The width, the position of the quasiparticle pole and the effect of multiparticle contributions can
be adequately studied with the presented algorithm.
\par
\begin{figure}
\begin{minipage}{0.48\textwidth}
\centerline{\psfrag{DOT}{\tiny  $\hat{T}_c/m$}
\psfrag{GOT}{\tiny $T_c/m$}
\includegraphics[width=.85\textwidth]{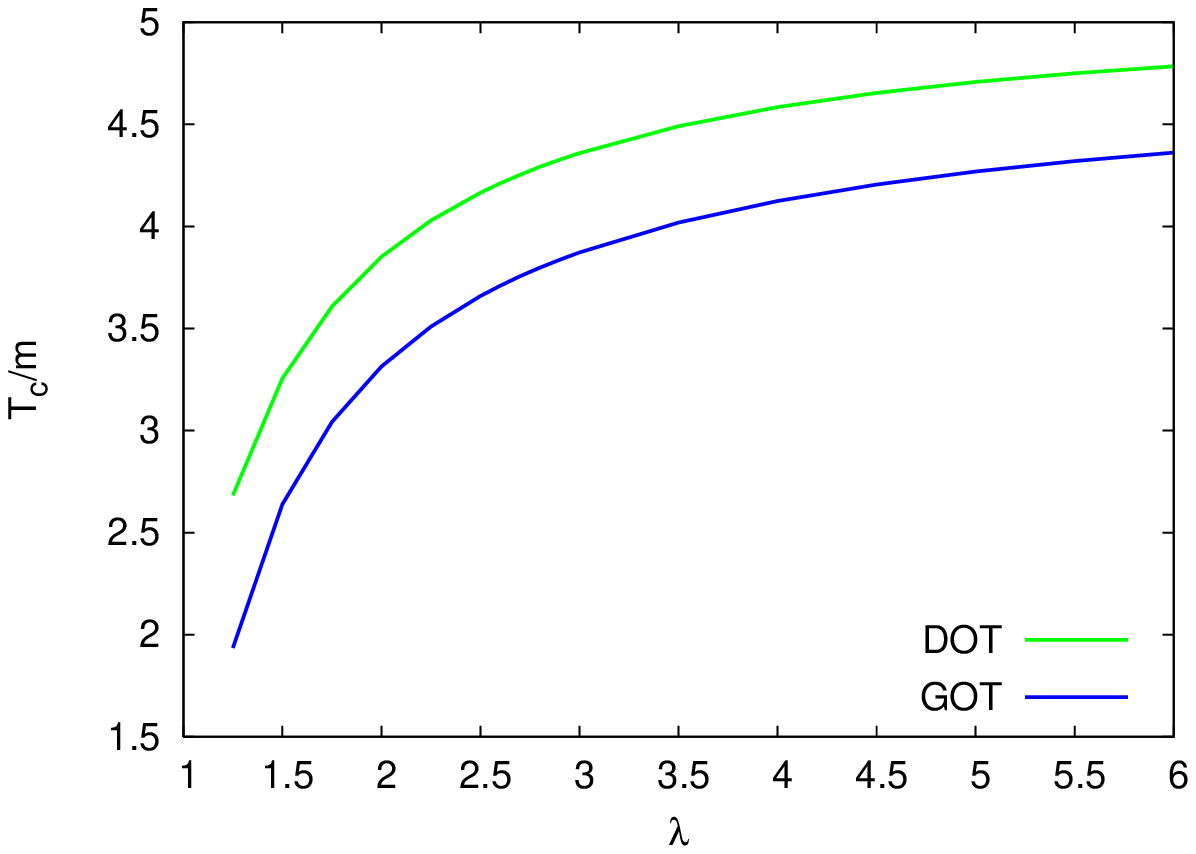}}
\end{minipage}
\begin{minipage}{0.48\textwidth}
\centerline{\psfrag{POT}{\tiny  $\hat{T}_c/m$}
\psfrag{ROT}{\tiny $T_c/m$}
\includegraphics[width=.85\textwidth]{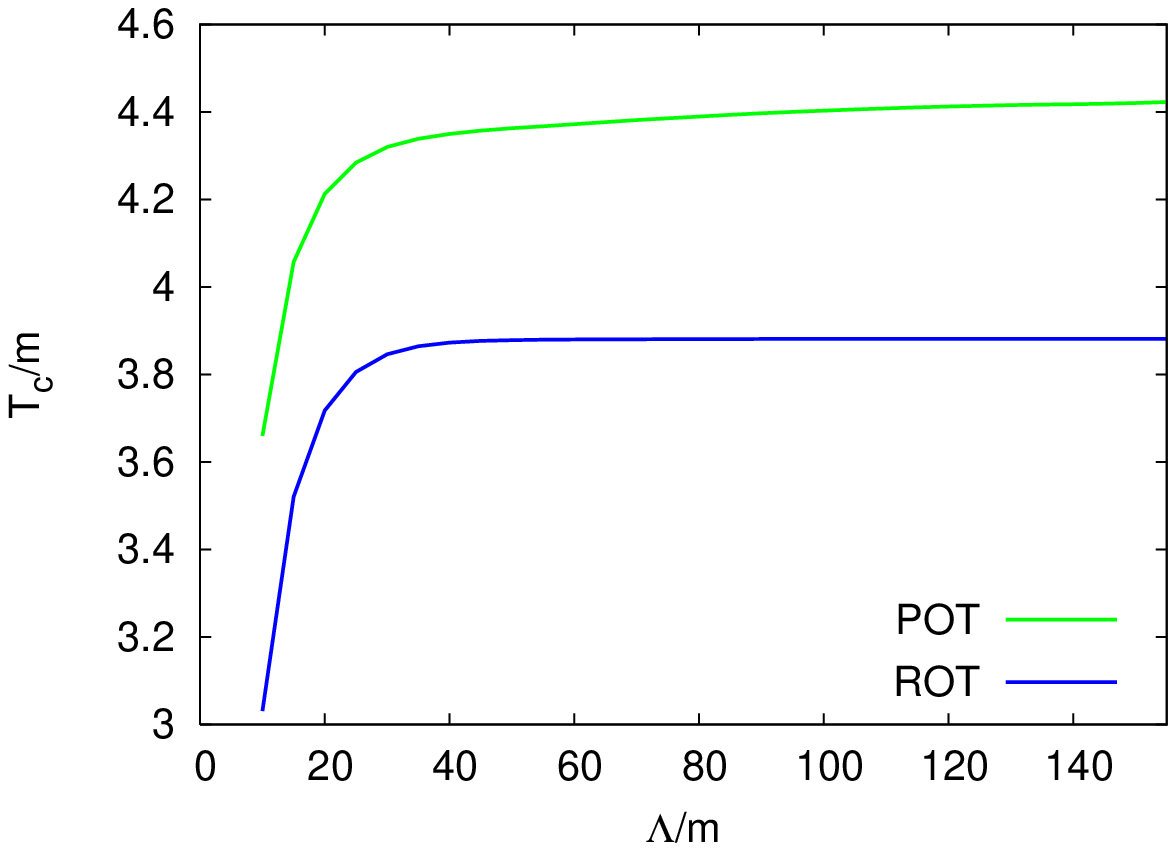}}
\end{minipage}
\vspace{-.75cm}
\label{fig:criticalT}
\caption{Left: Critical temperatures for several couplings and $T_\star/m=5$. Right: UV cut-off dependence for
$\lambda=3$.}
\vspace{-.75cm}
\end{figure}

We can also look at the critical behaviour in a system with broken symmetry. A simple quantity to
compute is the critical temperature which can be read off the 2-point functions at vanishing effective mass $M$. As we
discussed previously there are two possible 2-point functions ($\Gs$ and $\hat{G}$), for which the corresponding
effective masses are (in
the symmetric phase)
\begin{align}
M^2(T,\lambda)&=m^2+\Sigma^l(T,\lambda),\\
\hat{M}(T,\lambda)&=\hat{G}^{-1}(p=0,T,\lambda)=\frac{\delta^2 \Gamma[\phi,\Gs[\phi]]}{\delta
\phi^2}\Big|_{\phis}(p=0,T,\lambda).
\end{align}
Starting from $T_\star$ and reducing the temperature, the critical values $T_c$ and $\hat{T}_c$ are
found when $M^2(T_c,\lambda)$ and $\hat{M}(\hat{T}_c,\lambda)$ vanish (see Fig.~\ref{fig:criticalT}, left). We can also
vary the UV-cutoff in the calculation of the critical temperatures and therefore check the validity of the renormalization
procedure (see Fig.~\ref{fig:criticalT}, right). We find that both quadratic and logarithmic divergences (present respectively in
$T_c$ and $\hat{T}_c$) are properly renormalized and hence the renormalization is satisfactory.
\par
\begin{figure}
\begin{minipage}{0.48\textwidth}
\centerline{\includegraphics[width=.85\textwidth]{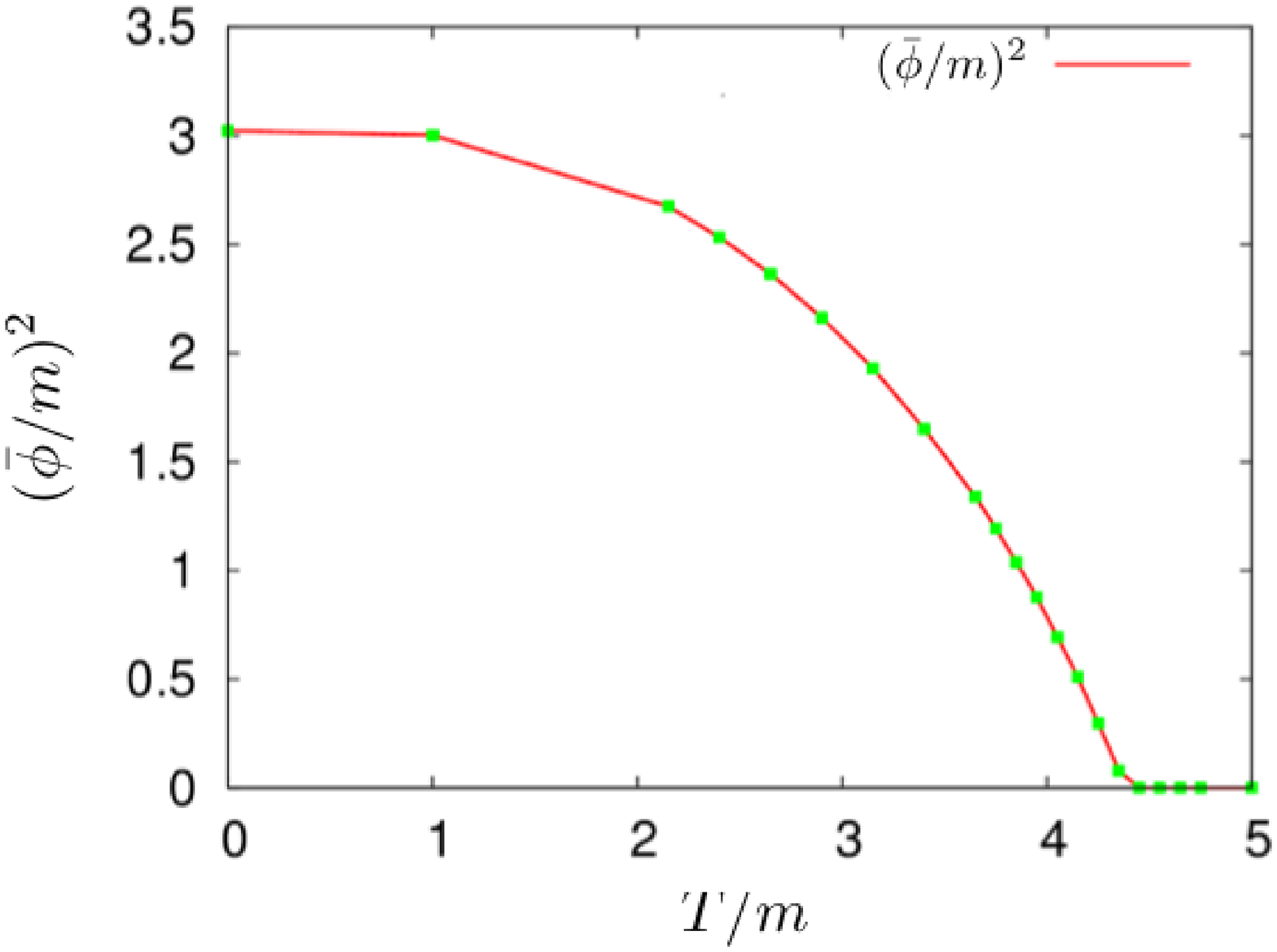}}
\end{minipage}
\begin{minipage}{0.48\textwidth}
\centerline{\includegraphics[width=.85\textwidth]{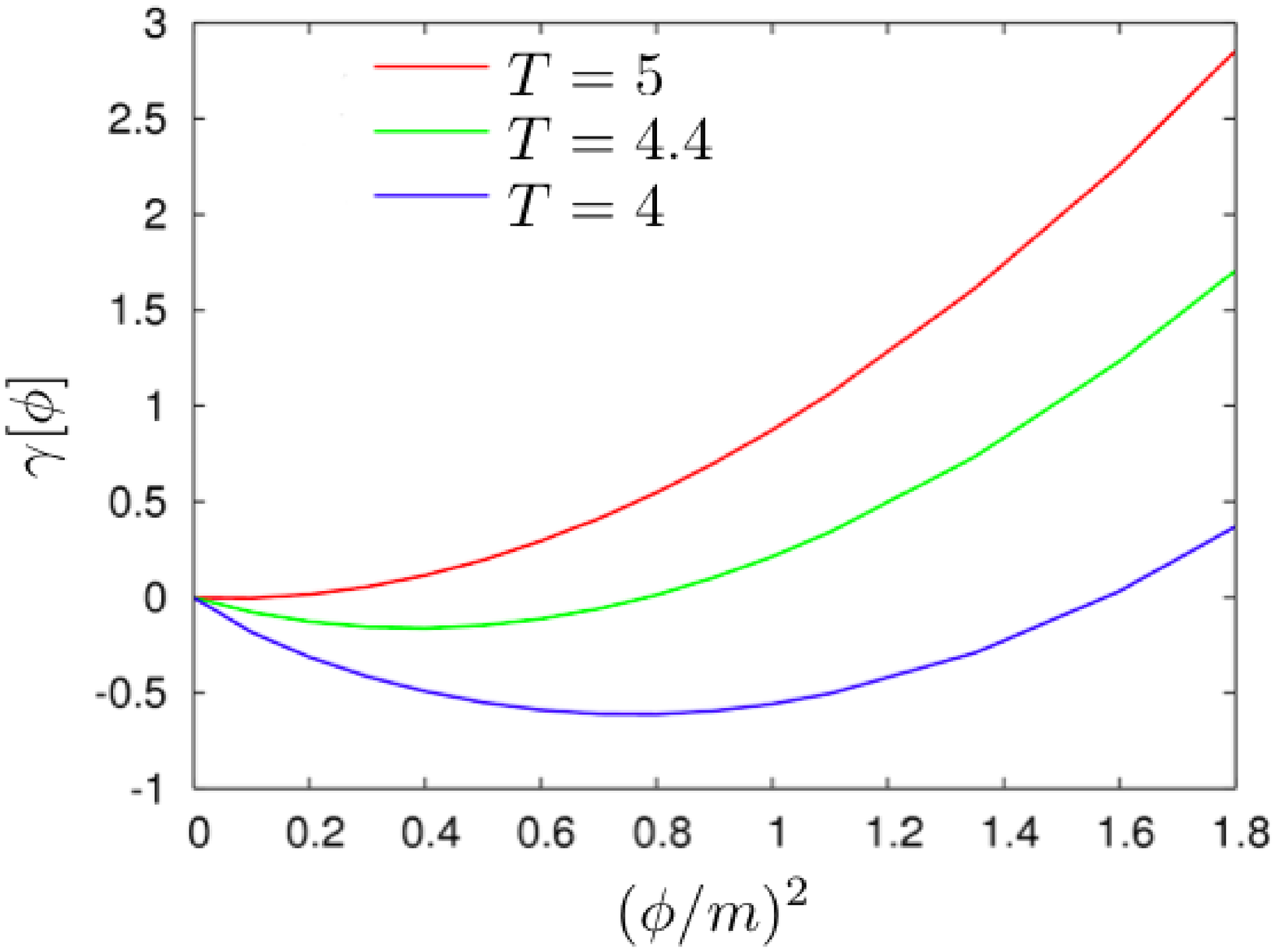}}
\label{fig:effectivepot}
\end{minipage}
\vspace{-.75cm}
\caption{Left: field expectation value $\phis^2$ as a function of $T$ for $\lambda=3$ and $T_\star/m=5$. Right:
Effective potential.}
\vspace{-.75cm}
\end{figure}

Solving in addition the field equation (\ref{eq:field}) allows us to study the phase transition (see
Fig.~\ref{fig:effectivepot}, left). We see the transition is clearly of second order. A similar conclusion is reached by
computing directly the effective potential $\gamma[\phi]$ (see Fig.~\ref{fig:effectivepot}, left). We checked 
that $\gamma[\phis]$, from which a renormalized free energy and pressure can be extracted, is cutoff independent.


\end{document}